\documentclass{raa_rb}          

\usepackage{graphicx,times,epsf}             %for PS/EPS graphics inclusion, new
\usepackage{amsmath}
\usepackage{makecell}
\usepackage[utf8]{inputenc}
\usepackage{multirow}
\newcommand{\Msold}{M$_{\odot}$\,yr$^{-1}$}

\begin{document}
   \title{Morphology and kinematics of the gas envelope of the variable AGB star $\pi^{1}$ Gruis 
\thanks{This paper makes use of the following ALMA data: \mbox{ADS/JAO.ALMA$\#<$2012.1.00524.S$>$}. 
ALMA is a partnership of ESO (representing its member states), NSF (USA) and NINS (Japan), 
together with NRC (Canada) and NSC and ASIAA (Taiwan), and KASI (Republic of Korea) in cooperation with the Republic of Chile. 
The Joint ALMA Observatory is operated by ESO, AUI/NRAO and NAOJ. The data are retrieved 
from the JVO portal (http://jvo.nao.ac.jp/portal) operated by the NAOJ}}
%   \subtitle{I. Place Your Subtitle Here}
   \volnopage{Vol.0 (200x) No.0, 000--000}      %%preserved for Editor. DOn't remove!
   \setcounter{page}{1}          %%starting page, preserved for Editor. DOn't remove!  
   \author{ P. T. Nhung\inst{}
          \and
           D. T. Hoai\inst{}
          \and
           P. N. Diep\inst{}
          \and
          N. T. Phuong\inst{}          
          \and
          N. T. Thao\inst{}
          \and
          P. Tuan-Anh\inst{}
          \and         
          P. Darriulat\inst{}
     }     
   \institute{Department of Astrophysics, Vietnam National Satellite Center, VAST, 18 Hoang Quoc Viet, Cau Giay, Hanoi, Vietnam; {\it pttnhung@vnsc.org.vn}}
   \date{}
   \abstract{
     Observations of the \mbox{$^{12}$CO(3-2)} emission of the circumstellar envelope (CSE) of the
     variable star $\pi^1$ Gru using the compact array (ACA) of the ALMA observatory have been
     recently made accessible to the public. An analysis of the morphology and kinematics of
     the CSE is presented with a result very similar to that obtained earlier for \mbox{$^{12}$CO(2-1)}
     emission by Chiu et al. (2006) using the Sub-Millimeter Array. A quantitative comparison
     is made using their flared disk model. A new model is presented that provides a
     significantly better description of the data, using radial winds and smooth evolutions of
     the radio emission and wind velocity from the stellar equator to the poles.
   }
   \authorrunning{P. T. Nhung et al.}            %author_head in even pages
   \titlerunning{}  % title_head in odd pages
   \maketitle
%% The author head (on even pages) and the title head (on odd pages) will be
%% automatically extracted from \author{} and \title{}. Whenever the title is too long,
%% you will be asked to supply a shorter one by inserting either \authorrunning{} or
%% \titlerunning{} before \maketitle. Anyway, you can specify your own heads.
%%
%%
%% Note: In the following text body of your manuscript, please note several differences from
%%       other major journals:
%% (1) \subsection{Please Capitalize the First Letter of Each Notional Word in Subsection Title}
%% (2) Please Capitalize the First Letter of Each Notional Word in all tables' captions

%
%________________________________________________ sections below

\section{Introduction}

$\pi^1$ Gruis (HIP 110478) is an SRb variable star of spectral type S5,7, namely in the
transition between being oxygen and carbon dominated (\cite{Pic1980}, \cite{ScaRos1976}, \cite{VanEck2011}), 
with a pulsation period of 195 days (\cite{PM2005}), a mean
luminosity of $\sim7240$ solar luminosities, an effective temperature of $\sim3100$ K (\cite{VanEck1998}) 
and technetium lines in its spectrum (Jorissen et al. 1993). According to Mayer et al. (2014), 
its original mass was around two solar masses but has now dropped to 1.5
only. It is close to the Sun, only 163 pc (\cite{VanLee2007}) and very bright, making it
one the best known S stars.

It has a companion at a projected angular distance of $2.8''$, meaning $\sim460$ AU, a
position angle of $200^\circ$ (\cite{Pro1981}) with an apparent visual magnitude of 6.55
(\cite{Ducati2002}, \cite{Ake1992}) and an orbital period of over 6200 years (\cite{Mayer2014}). 
Its large scale environment has been recently observed by Mayer et al. (2014)
at 70 $\mu$m and 160 $\mu$m using Herschel/PACS images and interferometric observations from the VLTI/AMBER archive. 
They observe an arc east of the star at a projection
distance of $\sim38''$, suggesting interaction with the companion.

Single dish observations of the CSE have been made by Knapp et al. (1999) using
the 10-meter telescope of the Caltech Submillimeter Observatory in CO(2-1) and by
Winters et al. (2003) at ESO with the SEST 15-m antenna in both CO(1-0) and CO(2-1).
The line profiles display extended wings with a double horned structure at low velocity.
The mass-loss rate is measured to be $\sim2.7\times10^{-6}$ \Msold, with an expansion
velocity of 14.5 kms$^{-1}$ (\cite{GB2008}) and a dust to gas ratio of 380
(\cite{Groenewegen1998}); Sahai (1992) was first to map the CSE using the 15
meter Swedish-ESO Sub-millimeter telescope, scanning over the star in steps of half a
beam, and to give direct evidence for a fast bipolar outflow.

High spatial resolution observations of the CO(2-1) emission were recently made
by Chiu et al. (2006), here referred to as C2006, using the Sub-Millimeter Array,
producing channel maps and position-velocity diagrams interpreted as resulting from a
slowly expanding equatorial disk and a fast bipolar outflow. They show that the
expansion at low velocities ($<$15 kms$^{-1}$ ) is elongated east-west and at high velocities ($>$25
kms$^{-1}$ ) is displaced north (red-shifted) and south (blue-shifted). They propose a model
describing the low velocity component as a flared disk with a central cavity of radius 200
AU and an expansion velocity of 11 kms$^{-1}$ , inclined by $55^\circ$ with respect to the line of
sight; and the high velocity component as a bipolar outflow that emerges perpendicular to
the disk with a velocity of up to $\sim45$ kms$^{-1}$.

Recently, ALMA observations of the CO(3-2) emission of the circumstellar
envelope, using the ALMA compact array ACA, have been made publicly available
(ALMA01004143). With a spatial resolution comparable with that obtained in the lower
velocity range by C2006, they usefully complement their CO(2-1) observations. The
present work presents an analysis and modeling of these new data.

\section{Observations, data reduction and data selection}
\label{sec:Observations}

The star was observed on October 8$^{\text{th}}$, 2013, for six minutes on the line using the compact
array (ACA, seven 7-m antennae) configuration with beam size and position angle of
$4.45''\times2.48''$ and $-84^\circ$. The data have been reduced by the ALMA staff. The line Doppler
velocity spectra are available in sixty bins of 2 kms$^{-1}$ between $-60$ kms$^{-1}$ and 60 kms$^{-1}$
after subtraction of a local standard of rest Doppler velocity of $-11$ kms$^{-1}$ ; the measured
value given by Mayer et al. (2014) is $-11.7\pm2.5$ kms$^{-1}$ but we round it up to have the
origin between bins 30 and 31. Figure 1 (left) displays the line profile, with a double-horn
structure at low velocities and wings extending to $-36$ kms$^{-1}$ and 40 kms$^{-1}$ and a mean
value of $-0.4$ kms$^{-1}$. In what follows we restrict the analysis to this velocity interval. Both
horns are of similar intensities in the present data while the blue-shifted horn dominates in
the C2006 data and the red-shifted horn in the data of Knapp et al. (1999). Figure 1 (right) displays 
the distribution of the number of pixels and velocity bins over the measured flux
density. The peak associated with the effective noise has an rms value of 6.2 mJy arcsec$^{-2}$.

%%%%%%% fig 1 %%%%%%%%%%%%%
\begin{figure*}[!htbp]
\begin{center}
\includegraphics[scale=0.24]{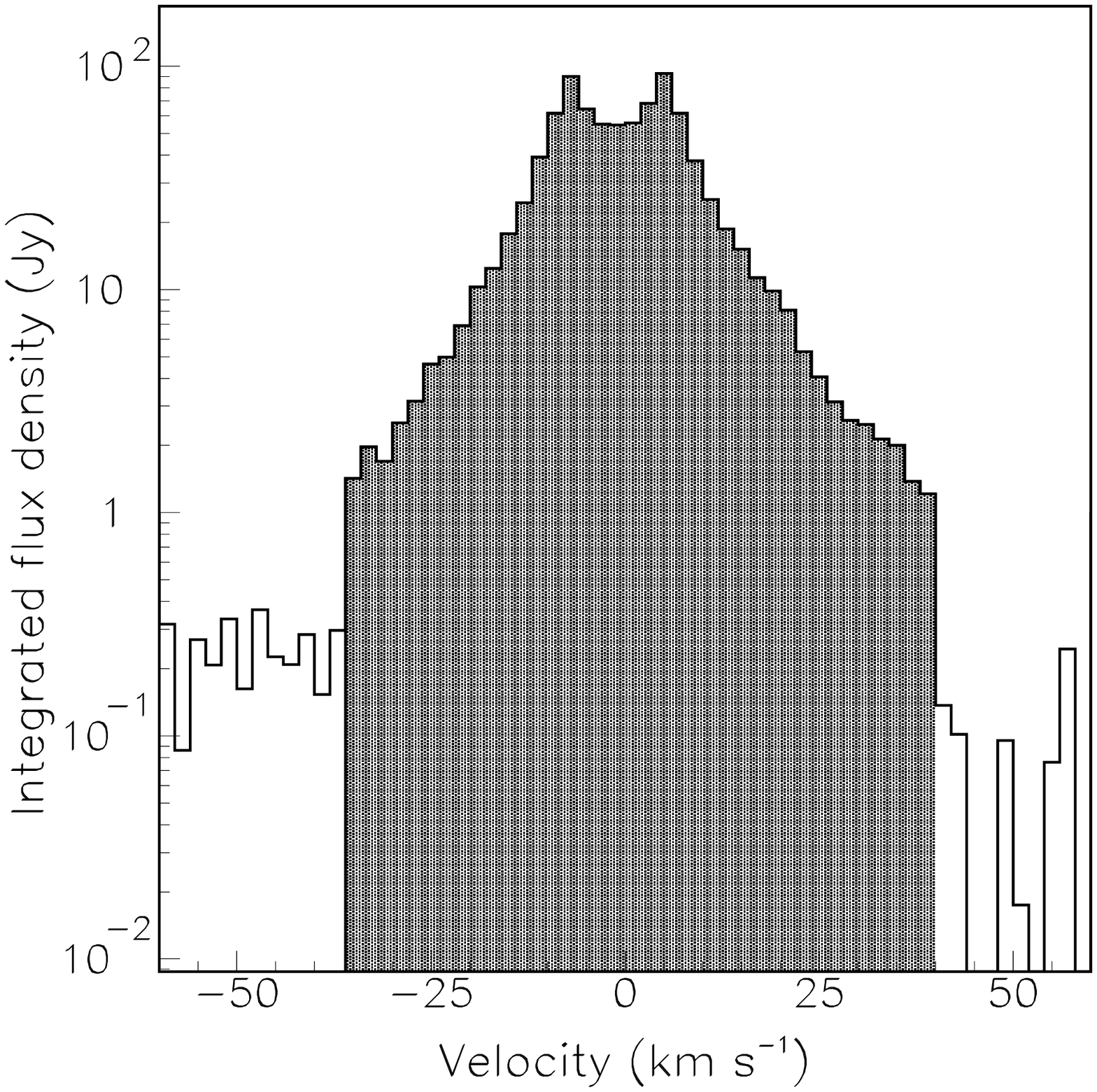}
\includegraphics[scale=0.24]{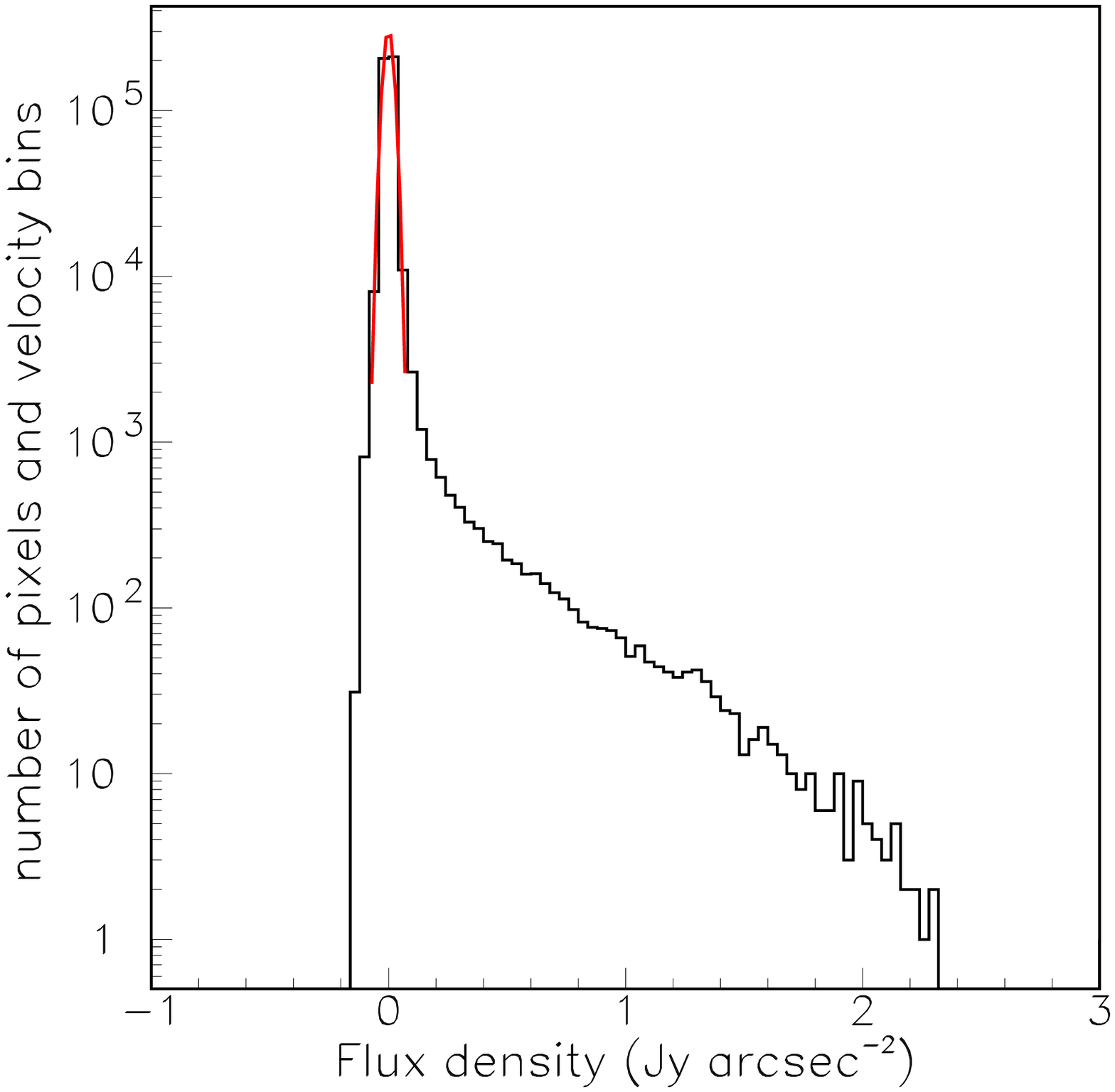}
\caption{Left: Integrated flux densities (Jy) as a function of Doppler velocities (kms$^{-1}$). Right:
Distribution of the number of pixels and velocity bins as a function of measured flux density in
\mbox{Jy arcsec$^{-2}$}. The line is a Gaussian fit centred at the origin with an rms value 6.2 mJy arcsec$^{-2}$.}\label{Fig1}
\end{center}
\end{figure*}

Figure 2 displays the $y$ and $z$ distributions ($y$ points east and $z$ points north) centred
on the peak of maximal emission taken as origin of coordinates: pixels are $0.5''\times0.5''$ and the mean values after centering are $0.28''$ and
$0.07''$ respectively, smaller than the pixel size. Also shown is the sky map of the flux integrated over Doppler
velocity and multiplied by $R=\sqrt{y^2 +z^2}$ in Jy arcsec$^{-1}$ kms$^{-1}$ in order to better reveal
inhomogeneities at large distances (a $r^{-2}$ dependence of the flux density means a $R^{-1}$ dependence of the integrated flux). In what follows, we limit the analysis to $R<7''$ in order to keep the
effects of noise and side lobes at a low level and apply a 3$\sigma$ cut to the measured flux densities (0.02 Jy arcsec$^{-2}$). 
From Mamon et al. (1988) we estimate the
effective radius of dissociation of CO molecules by the interstellar UV radiation to be
$\sim6700$ AU, namely $\sim41''$.

%%%%%%% fig 2 %%%%%%%%%%%%%
\begin{figure*}[!htbp]
\begin{center}
\includegraphics[scale=0.2]{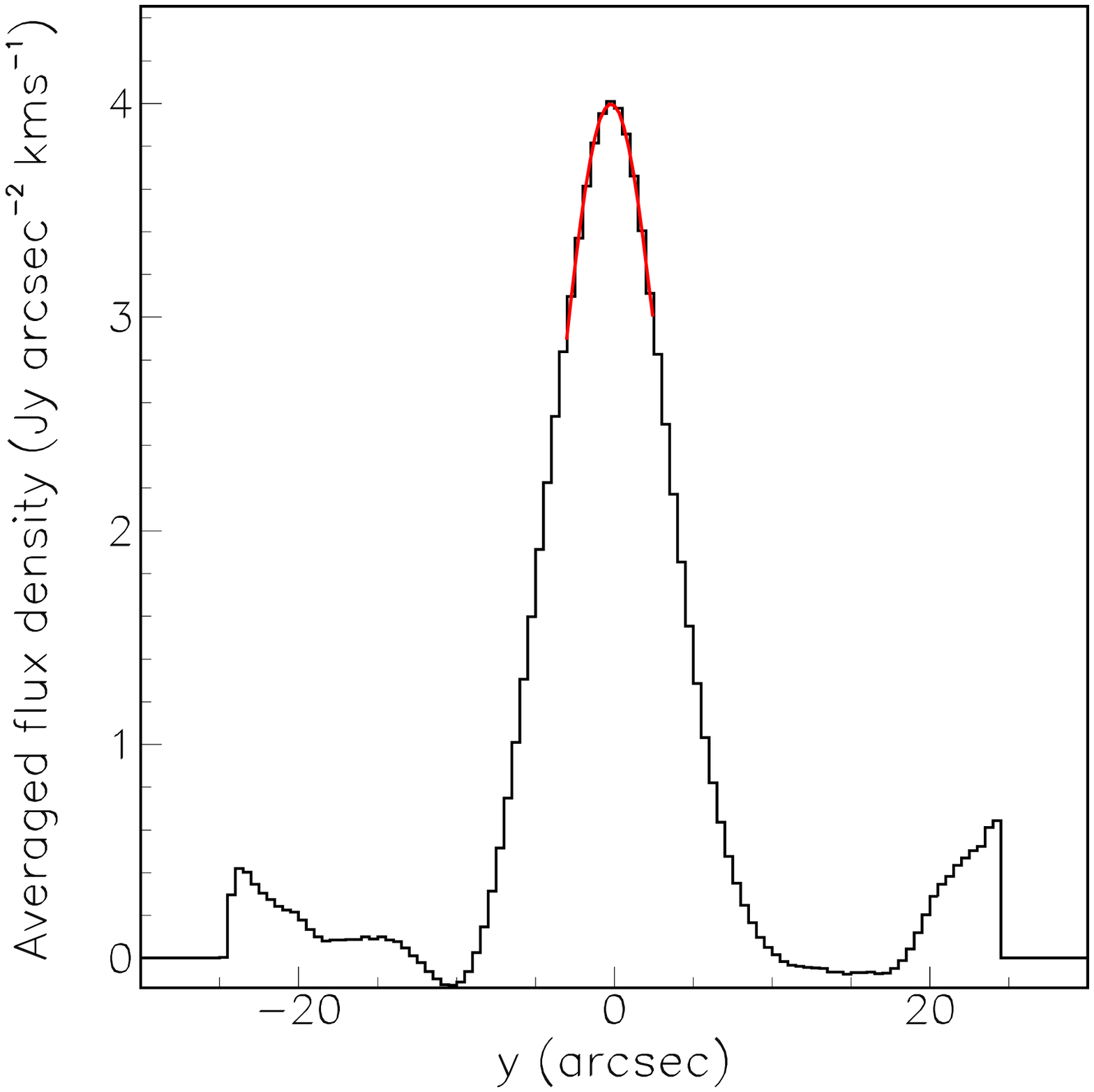}
\includegraphics[scale=0.2]{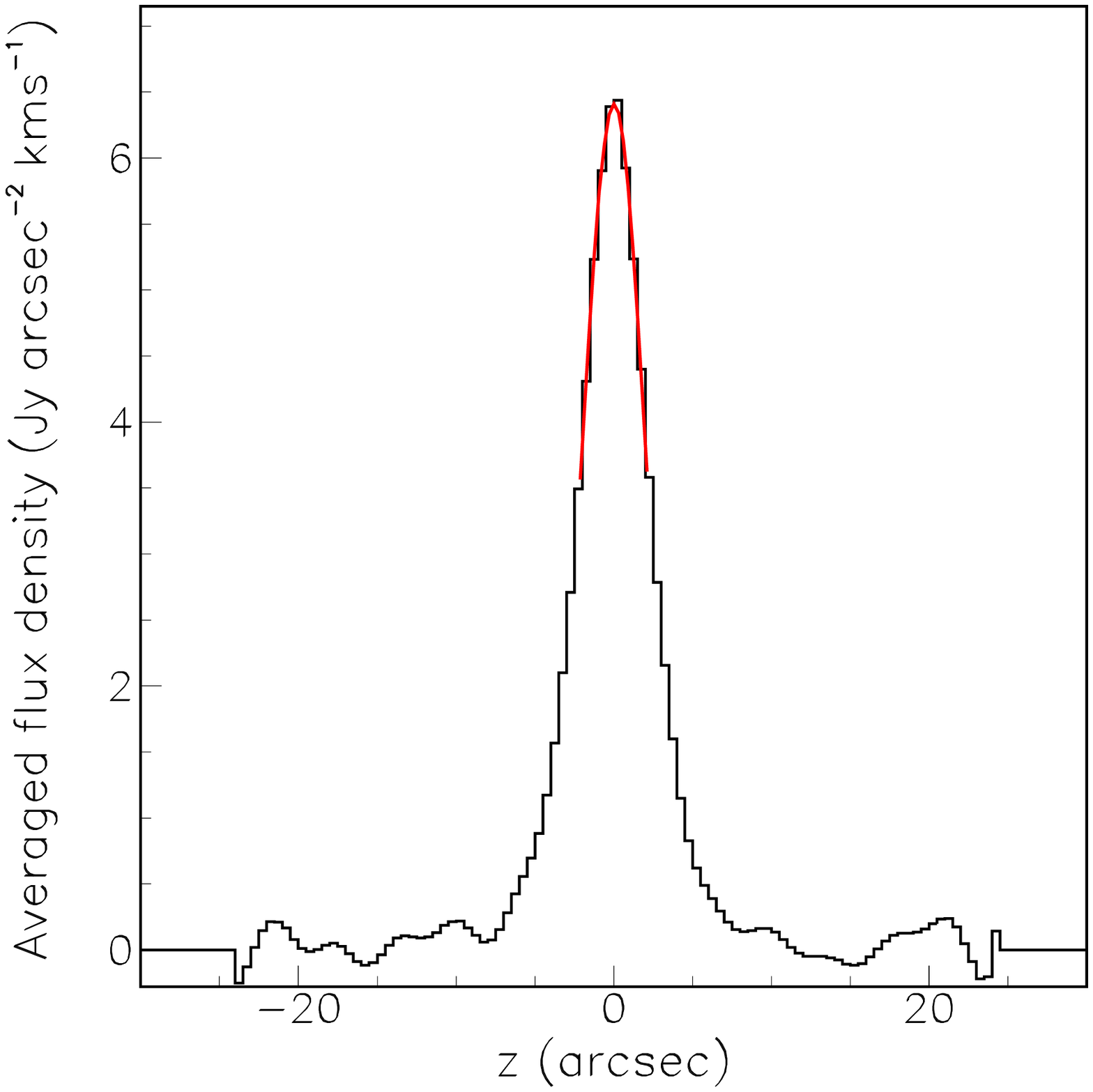}
\includegraphics[scale=0.3]{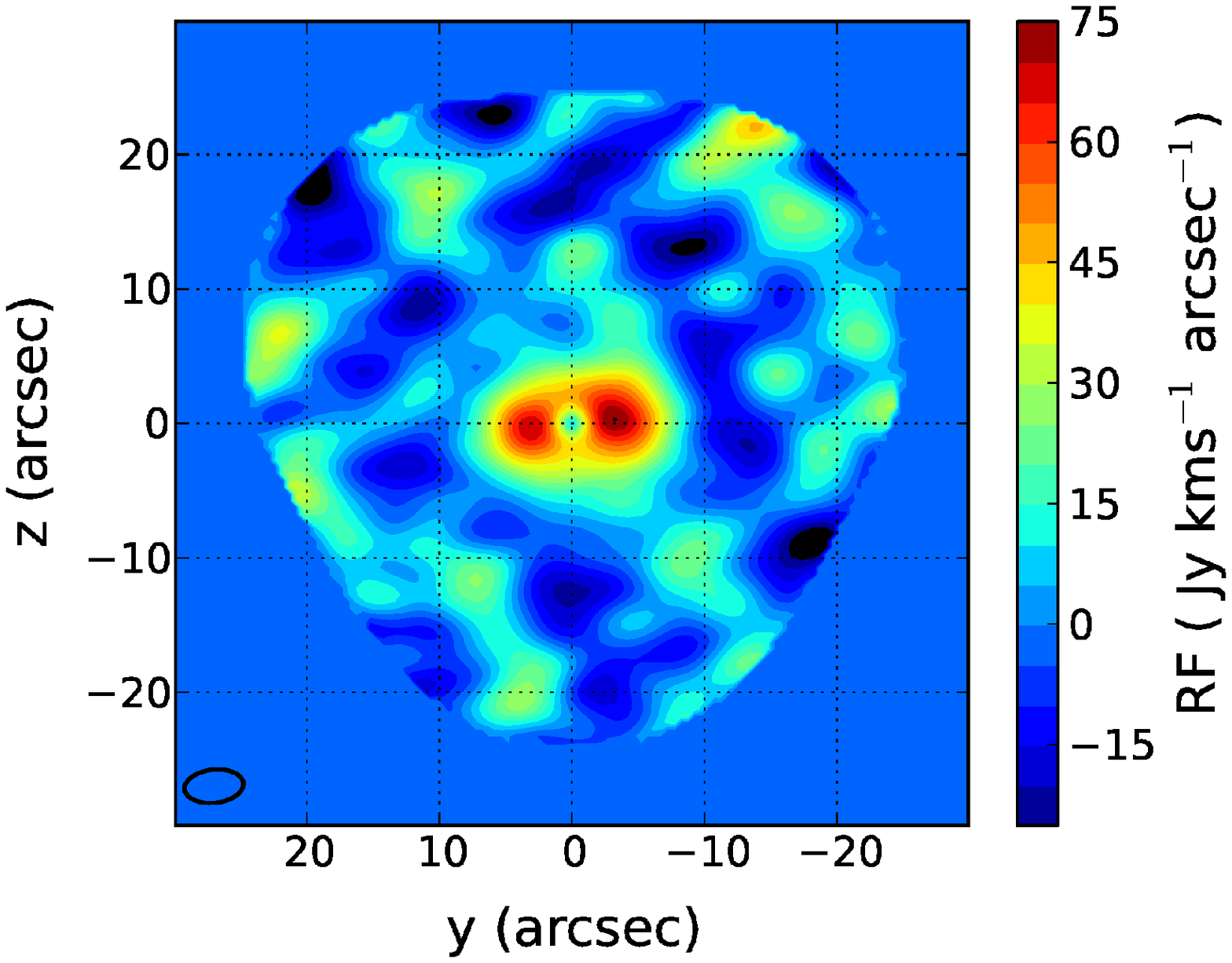}
\caption{Angular projected distances $y$ (left) and $z$ (middle) distributions for pixels where data
are available and for the Doppler velocity in the [$-36$ kms$^{-1}$ , 40 kms$^{-1}$] interval. The curves are
Gaussian fits having mean values of $0.28''$ and $0.07''$ respectively. Their sigma values are 3.38$''$ and 1.98$''$ respectively. Right: sky map of the flux integrated over Doppler velocity and multiplied by $R=\sqrt{y^2 +z^2}$ in Jy kms$^{-1}$ arcsec$^{-1}$ (this trivially produces a dip at the origin). The beam is shown in the lower left corner.}\label{Fig2}
\end{center}
\end{figure*}

\section{Morphology and kinematics of the observed CSE}
\label{sec:Morphology and kinematics}

Figure 3 displays channel maps and Figure 4 (left) maps the mean Doppler velocity over
the plane of the sky. The data are qualitatively similar to those published by C2006 for
CO(2-1). 
%%%%%%% fig 3 %%%%%%%%%%%%%
\begin{figure*}[!htb]
\begin{center}
\includegraphics[scale=0.37]{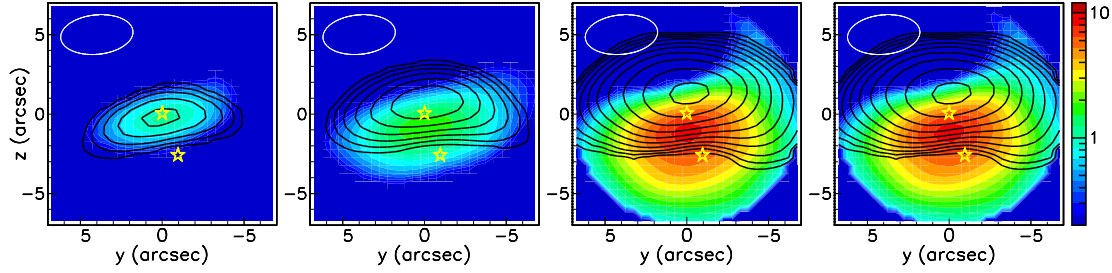}
\caption{Sky maps ($y$ and $z$ in arcseconds) of the integrated flux (Jy kms$^{-1}$ arcsec$^{-2}$) in Doppler velocity intervals of,
in absolute value and from left to right, 27 to 36 kms$^{-1}$, 18 to 27 kms$^{-1}$, 9 to 18 kms$^{-1}$ and 0 to 9
kms$^{-1}$. Colored maps are for negative velocities and contours for positive velocities in a same
interval of $|V_x|$. A same logarithmic scale is used for all panels. Stars indicate the position of the
primary and its companion. The beam is shown in the upper left corner of each panel.}\label{Fig3}
\end{center}
\end{figure*}
They give evidence for a north-south wind perpendicular to the east-west
elongation of the emission seen on the channel maps, as very clearly evidenced from their
dependences on position angle $\varphi=\tan^{-1}(z/y)+\pi$, measured clockwise from west (Figure 4,
middle and right), well described by simple sine wave fits of the form
$35.6+13.0\sin(2\varphi+101^\circ)$ (in Jy kms$^{-1}$ arcsec$^{-1}$) and $-0.49-6.92\sin(\varphi+7^\circ)$ (in kms$^{-1}$).
The mean Doppler velocity displayed in the right panel of the figure
is defined as $\sum{V_xf(R,\varphi,V_x)}/\sum{f(R,\varphi,V_x)}$ where the sums run over Doppler velocities 
and over pixels contained in a wedge corresponding to a given $\varphi$ bin.

The contribution of the present observations to our understanding of the physical
properties of the CSE of $\pi^1$ Gru is not sufficient to justify a detailed analysis of the kind
presented by C2006, which would essentially repeat what they have already done. 
We limit instead our discussion of the present data to the modelling of the morphology and
kinematics of the CSE using the concept of effective density, which we define as
$\rho(x,y,z)=f(y,z,V_x)\text{d}V_x/\text{d}x$. Here, $x$ is measured along the line of sight away from the
observer (positive $x$ values are red-shifted) and $f(y,z,V_x)$ is the measured flux density in
pixel $(y,z)$ at Doppler velocity $V_x$. With this definition, the integrated flux measured in
pixel $(y,z)$ is $F(y,z)=\int f(y,z,V_x)\text{d}V_x=\int\rho(x,y,z)\text{d}x$. The effective density is proportional to the
gas density but depends also on the temperature, the population of the emitting state and
the emission and absorption probabilities.
%%%%%%% fig 4 %%%%%%%%%%%%%
\begin{figure*}[!htb]
\begin{center}
\includegraphics[height=4.45cm, trim=2.cm .85cm 0.cm 0.cm,clip]{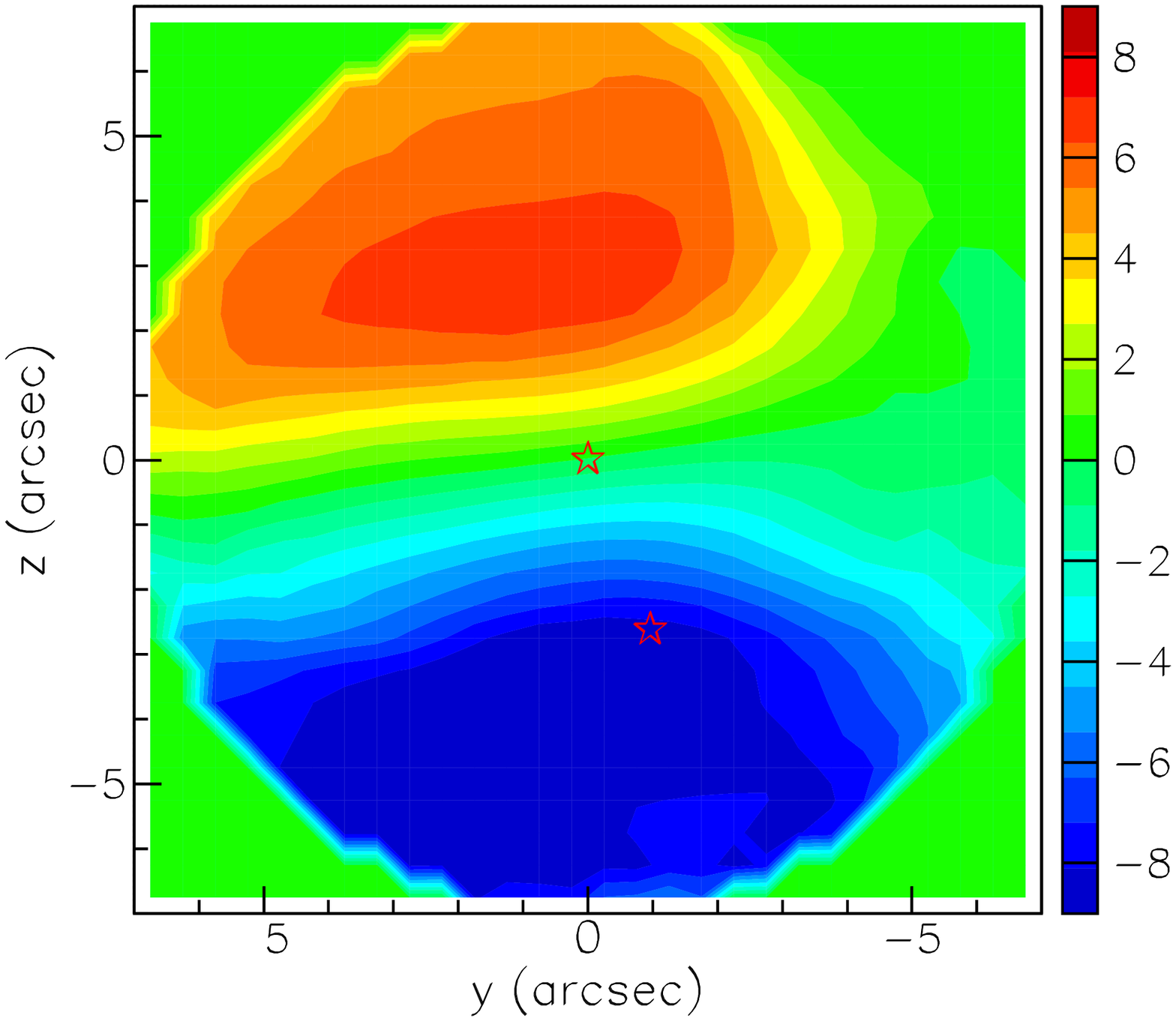}
\includegraphics[height=4.3cm]{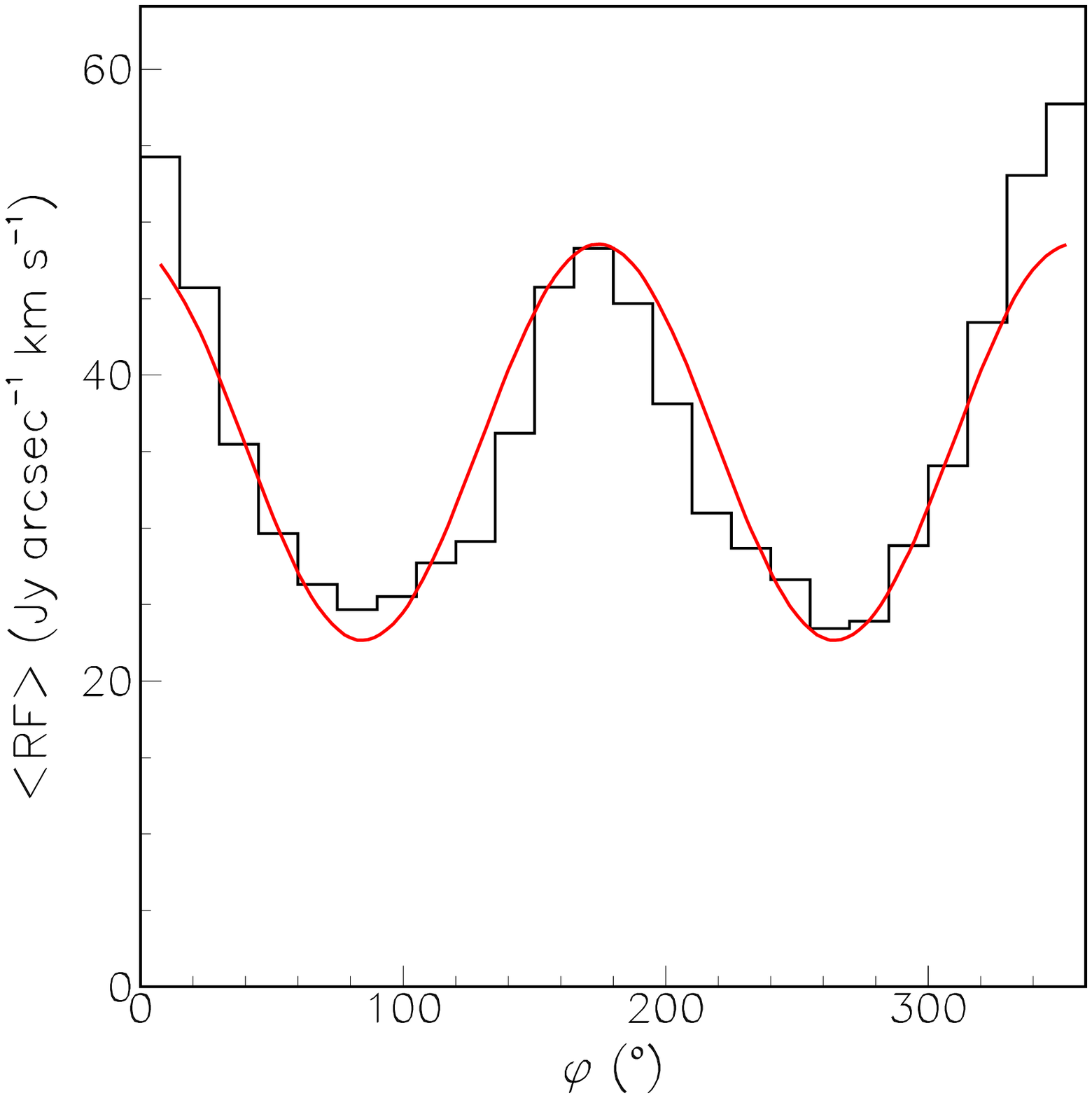}
\includegraphics[height=4.3cm]{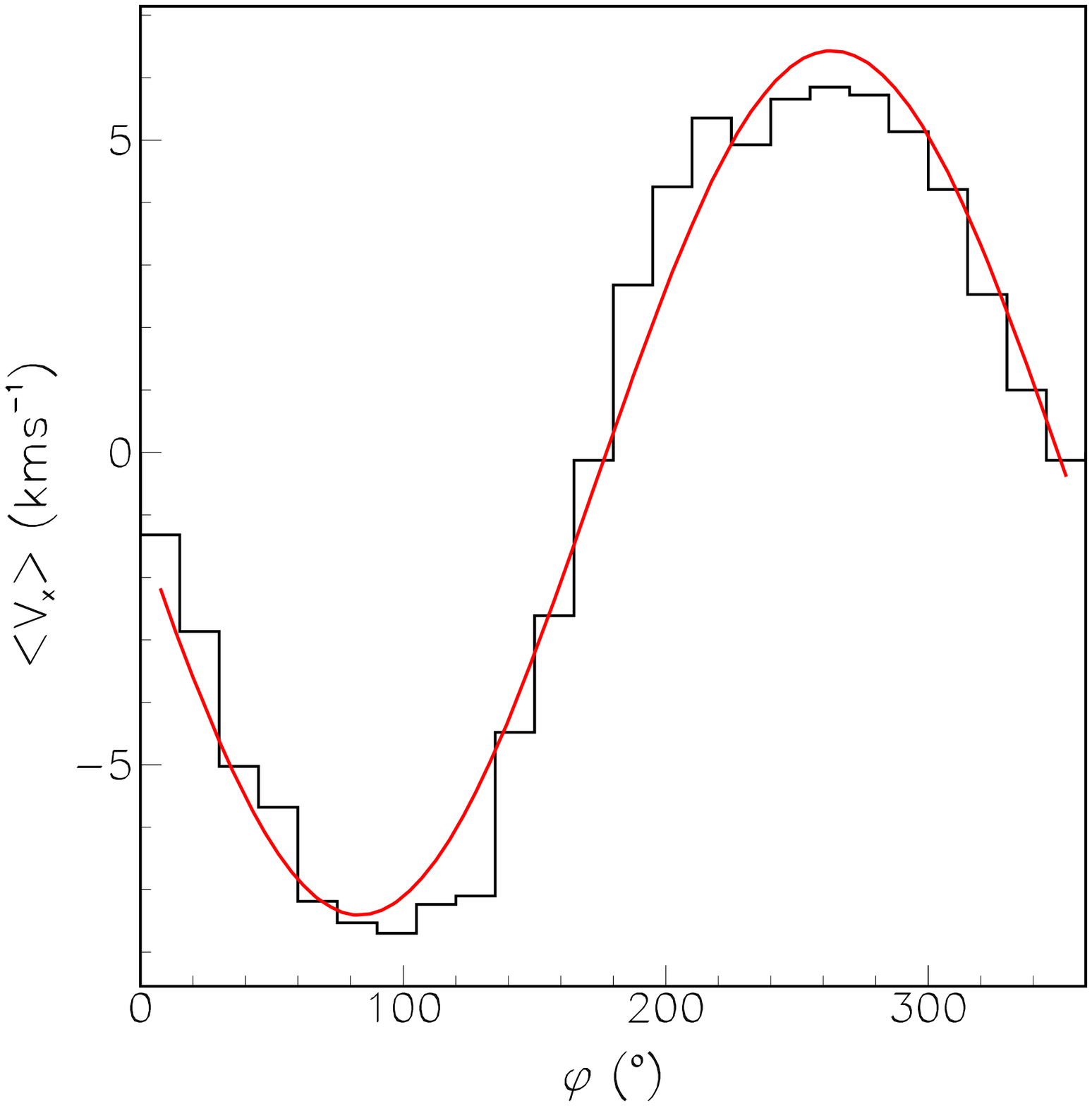}
\caption{Left: sky map of the mean Doppler velocity (in kms$^{-1}$).
Stars indicate the position of the primary and its companion. Middle: dependence on $\varphi$ of
the flux integrated over Doppler velocity, averaged over $R$ and multiplied by $R=\sqrt{(y^2+z^2}$ in 
\mbox{Jy arcsec$^{-1}$ kms$^{-1}$}. Right: dependence on $\varphi$ of the 
mean Doppler velocity (kms$^{-1}$) averaged over $R$.
The curves are sine wave fits of the form $35.6+13.0\sin(2\varphi+101^\circ)$ (middle) and 
\mbox{$-0.49-6.92\sin(\varphi+7^\circ)$} (right).}\label{Fig4}
\end{center}
\end{figure*}

%%%%%%% fig 5 %%%%%%%%%%%%%
\begin{figure*}[!htb]
\begin{center}
\includegraphics[scale=0.38]{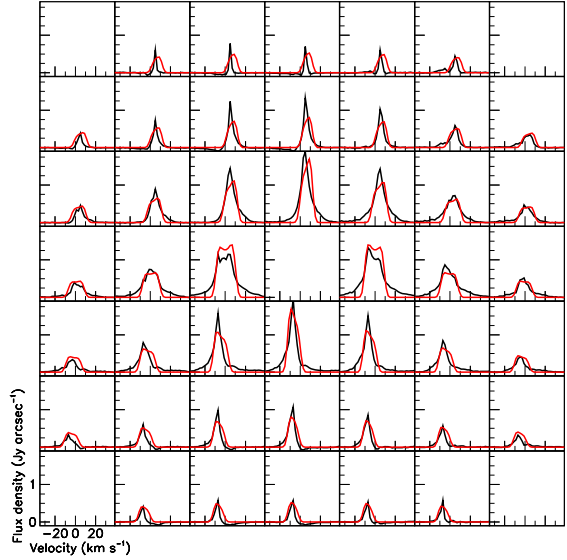}
\caption{Spectral map of the present ACA data restricted to the region $R<7''$. Each pixel is
$2''\times2''$. The red curves are the best fit using the C2006 model. The central pixel is excluded from
the fit.}\label{Fig5}
\end{center}
\end{figure*}

In the present data, the very high velocity winds, up to 80 kms$^{-1}$, hinted at by
Knapp et al. (1999) and C2006 would require longer observation times to be solidly
established. Yet, the winds observed here extend over some $\pm40$ kms$^{-1}$, which is quite a
large span, and join smoothly to the low velocity profile. As the high spatial resolution
C2006 data cover only $\pm25$ kms$^{-1}$, the contribution of the present data is significant in
this respect.

\section{Flared disk model of the CSE}
\label{sec:Flared disk}

In a first step, we reproduce the results obtained by C2006 for CO(2-1). Namely the
emission is confined to a flared disk having inner radius $r_{min}$, 
inner thickness $2z_{min}$ and a wedge angle $\pm\beta_{wedge}$. 
The orientation of its axis is defined by two angles, $\theta$ measuring its
inclination with respect to the line of sight and $\psi$ measuring the position angle of its
projection on the sky plane counter clockwise from north. We take the effective density
proportional to $r^{-n}$ and the wind velocity radial with constant intensity $v_{rad}$ . We use a pixel
size of $2''\times2''$ without convolving the model by the beam and integrate along the line of
sight up to $r=41''$, the mean UV dissociation radius. A Gaussian smearing of $\pm2$ kms$^{-1}$ is
included as in C2006. The values of the parameters giving the best fit are compared in
Table 1 with the values found by Chiu et al. for CO(2-1). They are very close to each
other. Note that the $r^{0.25}$ dependence of the flux of matter found by C2006 and the $r^{-0.7}$
dependence of the temperature that they use cannot be directly compared with the $r^{-2.03}$ dependence of the effective density used in our model, but both give adequate descriptions of the observations. 
The measured spectral map is compared with the model result in
Figure 5. As found earlier by C2006 for CO(2-1), the model is seen to give a good
qualitative description of the data but important deviations are observed. In particular the
high wind velocity wings are obviously not reproduced by the model: the value of $v_{rad}$ is
determined by the double-horned profile at low velocities and cannot produce Doppler
velocities in excess of $v_{rad}$.

%%%% Table 1 %%%%%%
\begin{table}
\caption[]{Results of the C2006 model fits for CO(2-1) and CO(3-2).}
\centering
\renewcommand{\arraystretch}{1.2}
\begin{tabular}{c|c|c}
\hline
  Parameter &  \makecell{C2006 \\CO(2-1)} & \makecell{This work \\ CO(3-2)}\\
  \hline
$\theta$  & $145^\circ$   & $140^\circ$  \\ % new variable
$\psi$    & $6^\circ$     &  $6^\circ$    \\
$V_{rad}$  & 11 kms$^{-1}$     &  12.5 kms$^{-1}$         \\
$z_{min}$  & $0.84''$     & $1.0''$   \\
$r_{min}$  & $1.31''$     & $1.3''$   \\
$\beta_{wedge}$  & $25^\circ$  &  $25^\circ$ \\
$n$       &  -  &  2.03  \\
\hline
\end{tabular}
\end{table}

\section{A new model of the CSE}
\label{sec:new model}
In order to improve on the C2006 model, one needs therefore to account for the presence
of higher wind velocities than obtained from this model. Figure 6 displays position-
velocity diagrams in the $V_x$ vs $\varphi$ plane for two different intervals of $R$ (the term is
normally used in Cartesian coordinates but the information carried by such diagrams is of
the same nature in polar coordinates). It shows continuity between low and high absolute
values of the Doppler velocity and between low and high values of $R$. This suggests a
smooth variation of the expansion velocity from the star equator, where the disk stands,
to the star poles on the star axis perpendicular to the disk rather than a bipolar jet
confined around the star axis. This had been already remarked by C2006 who proposed
an explanation in terms of “material punched out of the disk by the bipolar outflow”
and/or a “bipolar outflow simply having a very large opening angle”.

We construct therefore a new model allowing for continuity from equator to poles.
A first exploration of different parameterizations suggested allowing for the effective
density to extend to small values of $r$, and using a much smaller smearing of the Doppler
velocities than assumed in C2006. In addition, it showed that adding rotation about the
star axis does not bring significant improvement. On this basis, we write the radial
expansion velocity as $V_{pole}\exp(-0.5\cos^2\alpha/\sigma_{pole}^2)$ where $\alpha$ is the stellar latitude and the
effective density as $\rho=\rho_0 r^{-n}(\cos\alpha)^p[1-\exp(-0.5r^2/\sigma_{cut}^2)]$ namely five parameters to be
adjusted. In addition we include four additional parameters: the direction of the star axis,
measured by $\theta$ and $\psi$ as before, the smearing parameter $\sigma_v$ and a small shift $\delta v$ of the
origin of the Doppler velocity scale. The best fit value of $\chi^2$ is divided by 2.9 with respect
to the C2006 model fit and the corresponding values of the model parameters are listed in
Table 2. The value of $\rho_0$, obtained from the normalization between data and model, is
144 Jy arcsec$^{-3}$kms$^{-1}$. The fit to the spectral map is illustrated in Figure 7.

%%%%%%% fig 6 %%%%%%%%%%%%%
\begin{figure*}[!htbp]
\begin{center}
\includegraphics[scale=0.4]{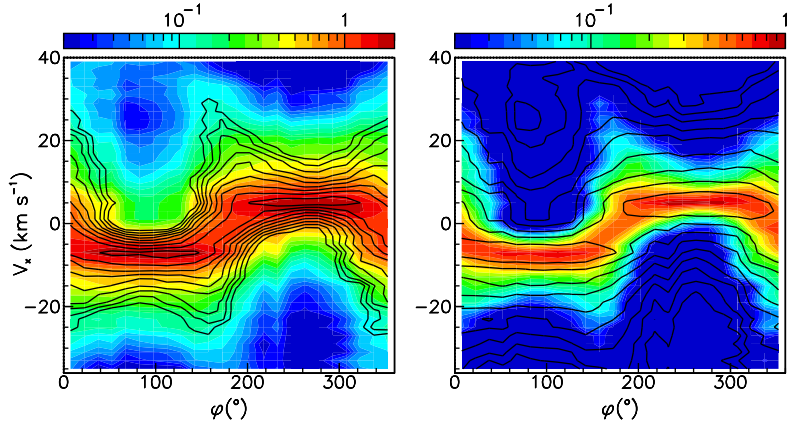}
\caption{Position-velocity diagrams in the Doppler velocity $V_x$ vs position angle $\varphi=\tan^{-1}(z/y)+\pi$
for $R<3''$ (left) and $3''<R<7''$ (right). Contours in the left (right) panel correspond to the right
(left) panel. The colour scale is in the unit of Jy arcsec$^{-2}$.}\label{Fig6}
\end{center}
\end{figure*}

%%%% Table 2 %%%%%%
\begin{table}
\begin{center}
\caption[]{Best fit values of the parameters in the new model. The quoted ``uncertainties'' list the
amounts by which a parameter has to be changed for $\chi^2$ to increase by 5\%.}

 \begin{tabular}{cccccccccc}
  \hline\noalign{\smallskip}
Parameter  & \makecell{$\theta$\\($^\circ$)} & \makecell{$\psi$\\($^\circ$)} & \makecell{$V_{pole}$\\(kms$^{-1}$)} &  \makecell{$\sigma_{pole}$\\(kms$^{-1}$)} 
& $n$ & $p$ & \makecell{$\sigma_{cut}$\\(arcsec)} & \makecell{$\sigma_v$\\(kms$^{-1}$)} & \makecell{$\delta v$ \\(kms$^{-1}$)}\\
  \hline\noalign{\smallskip}
Best fit value  & 140 & 6 & 64 & 0.55 & 3.51 & 4.1 & 3.3 & 0.2 & -1.2 \\ 
``Uncertainty''   & 2 & 5 & 3 & 0.08 & 0.12 &0.8 & 0.4 & $<1.2$ & 0.3 \\
  \noalign{\smallskip}\hline
\end{tabular}
\end{center}
\end{table}

%%%%%%% fig 7 %%%%%%%%%%%%%
\begin{figure*}[!htbp]
\begin{center}
\includegraphics[scale=0.38]{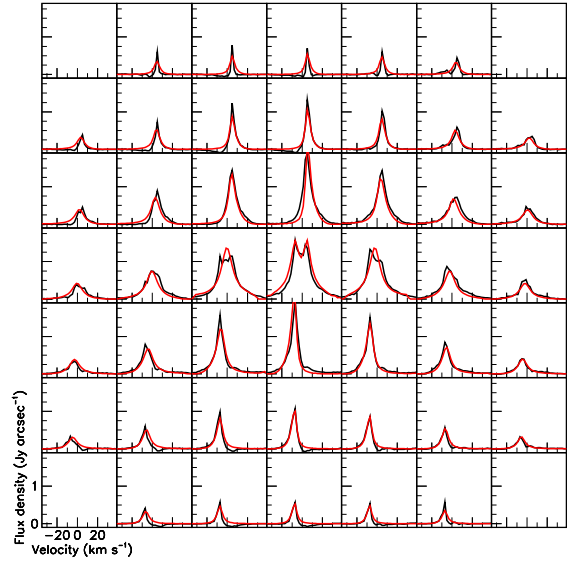}
\caption{Spectral map of the present ACA data restricted to the region $0.75''<R<7''$. Each pixel
is $2''\times2''$. The red curves are the best fit using the new model (Table 2).}\label{Fig7}
\end{center}
\end{figure*}

The dependence on stellar latitude of the radial expansion velocity and of the
effective density is illustrated in Figure 8. The relative crudeness of the present model
does not allow for a reliable quantitative evaluation of the uncertainties attached to the
model parameters. However, we can measure the sensitivity of the quality of the best fit,
i.e. the value of $\chi^2$, to the values taken by the model parameters as proportional to the
quantity by which they must be changed to increase the value of $\chi^2$ by $\sim5\%$. These values
are listed in Table 2. They ignore correlations between parameters, such as between $\rho_0$
and $\rho$. We checked that accounting for a velocity gradient does not significantly improve
the quality of the best fit. While the new model succeeds to account for the observed high
velocity wings, it underestimates the importance of the double horn structure (Figure 9).
Figure 10 displays sky maps of individual velocity bins in the horn region in an attempt
to improve the quality of the best fit. It shows that the horns populate low $y$ regions at
opposite $z$ values while the velocities in between the horns populate a low $z$ region with a
broader $y$ extension. We failed to find a simple modification of the new model that would
account for this feature.

%%%%%%% fig 8 %%%%%%%%%%%%%
\begin{figure*}[!htb]
\begin{center}
\includegraphics[scale=0.35]{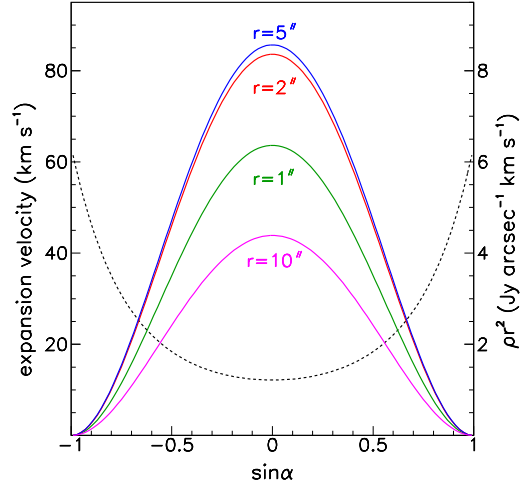}
\caption{Dependence on stellar latitude of the radial expansion velocity $V_{rad}$ and effective
density $\rho$ multiplied by $r^2$ corresponding to the best fit (Table 2). The effective density curves are
for $r=1''$, $2''$, $5''$ and $10''$ as labelled and their values can be read from the right-hand scale.}\label{Fig8}
\end{center}
\end{figure*}

%%%%%%% fig 9 %%%%%%%%%%%%%
\begin{figure*}[!htb]
\begin{center}
\includegraphics[scale=0.35]{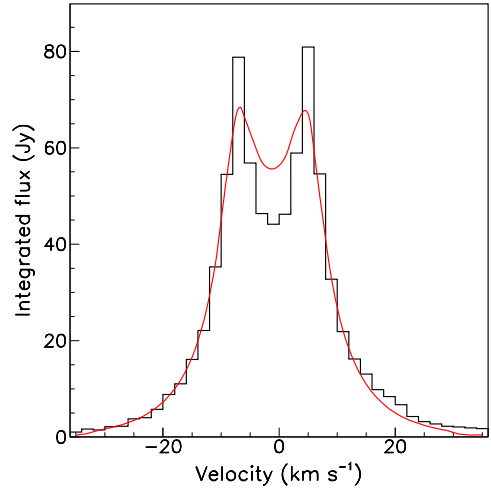}
\caption{Doppler velocity spectrum of the present ACA data restricted to the region $R<7''$ and
excluding the central $2''\times2''$ pixel. The red curve is the best fit using the new model.}\label{Fig9}
\end{center}
\end{figure*}

%%%%%%% fig 10 %%%%%%%%%%%%%
\begin{figure*}[!htb]
\begin{center}
\includegraphics[scale=0.35]{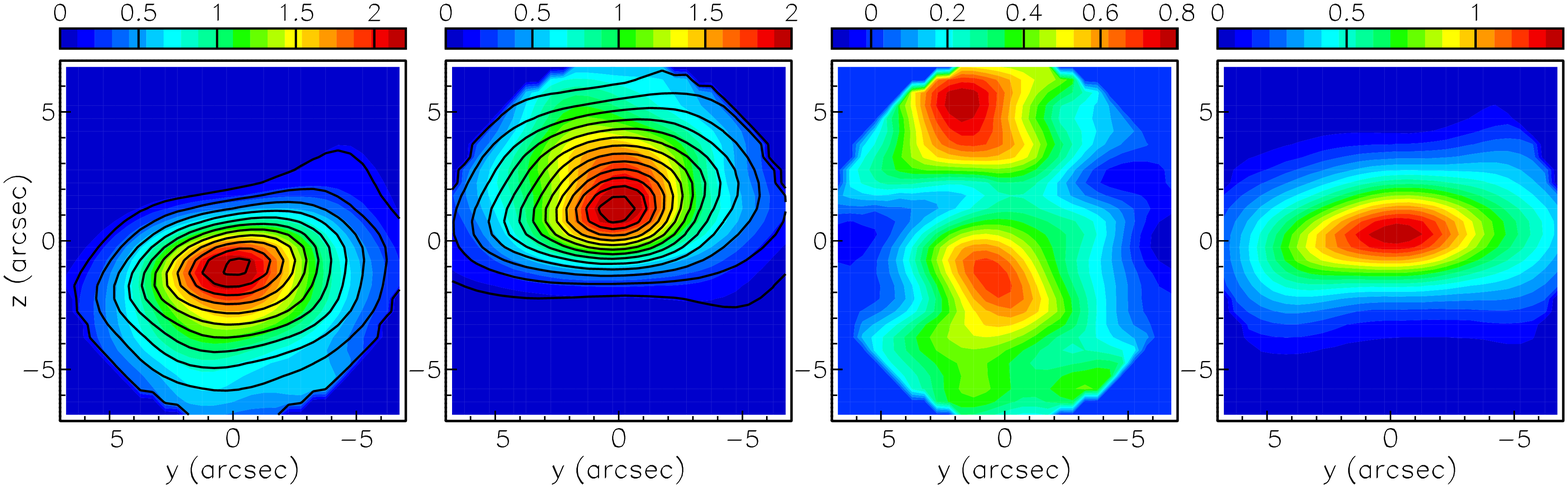}
\caption{Sky maps of the double-horned structure for selected Doppler velocity bins. Extreme
left: Doppler velocity bin of the blue-shifted horn. Contours show the map of the bins
immediately below and above the horn bin. Central left: Doppler velocity bin of the red-shifted
horn. Contours show the map of the bins immediately below and above the horn bin. Central
right: horn velocity bins with the average of the neighbour bins subtracted. The upper population
corresponds to the red-shifted horn, the lower population to the blue-shifted horn. Extreme right:
the average of the three central Doppler velocity bins between the horns. The colour scales are in the unit of Jy arcsec$^{-2}$.}\label{Fig10}
\end{center}
\end{figure*}

\section{Discussion and conclusion}
\label{sec:Discussion}

The new model brings a significant improvement over the flared disk model of C2006
and shows that there is no need for assuming the expansion to be confined to a flared disk
having sharp boundaries. A similar model has been used to describe AGB stars, such as
RS Cnc and EP Aqr, having a lower mass loss rate, at the scale of $10^{-7}$ rather than $10^{-6}$
\Msold~ (\cite{Hoai2014}, \cite{Nhung2015a}, \cite{Nhung2015b}).
However, in the present case, the effective density displays a strong oblateness while for
the lower mass stars it is nearly spherical. The model used here for $\pi^1$ Gru implies that the
effective density cancels at the poles (as $\cos^p\alpha$) but if we allow for it to take a positive
value $\rho_{pole}$  we find that its largest possible value making $\chi^2$ increase by at most $5\%$ is only
13 Jy arcsec$^{-3}$kms$^{-1}$ compared with 144 at the equator. We can therefore state that the
effective density drops by at least a factor 10 between equator and poles, compared with a
drop of less than $30\%$ for RS Cnc and EP Aqr. The present data do not allow for telling
apart the respective roles of actual gas density and temperature in such a strong latitudinal
dependence. The assumption made in C2006 of a purely radial temperature dependence is
probably unjustified in the presence of the strong winds observed here. Simultaneous
observation of at least two molecular lines would be necessary to disentangle the radial
from latitudinal dependences of the gas temperature and density.

As noted earlier by C2006, V Hya, a C star, shows striking similarities
with $\pi^1$ Gru. From the channel maps and position-velocity diagrams shown by Hirano et
al. (2004), it seems very likely that the present new model should give a good description
of the morphology and kinematics of the CSE of this star. This would suggest for these
AGB stars some continuity of the mass loss mechanism, with a mass loss rate evolving
from $\sim10^{-7}$ \Msold~ at an early stage (EP Aqr has no technetium in its
spectrum) to several $10^{-6}$ \Msold~ at the late stage of evolution of \mbox{V Hya} and
$\pi^1$ Gru. During this evolution, the effective density gets progressively depleted in the
polar regions and accordingly enhanced at the equator while the bipolar outflow is
present from the beginning with a radial expansion velocity of some 10 kms$^{-1}$ , increasing
to some 50 kms$^{-1}$ or more at the late stage of evolution. Moreover, the strong similarity
between EP Aqr and RS Cnc, a star having already experienced several dredge up
episodes, would suggest that this evolution is not linear in time but gets accelerated when
approaching the final stage. If such an interpretation were to be confirmed by future
observations, it would disfavour models assuming the presence of a thin equatorial gas
disk for this family of AGB stars. Indeed, direct evidence for such a thin gas disk has
never been obtained in the case of AGB stars, nor has evidence for rotation of the
equatorial region about the star axis, although some AGB stars, such as L$_2$ Pup, an
oxygen rich star, offer hints at the presence of a rotating disk (\cite{MII2013}, 
\cite{Kervella2014}, \cite{Lykou2015}).

It is often argued (for a recent review see \cite{LC2015}) that the
most likely mechanism for the formation of an enhanced equatorial density and of a 
bipolar outflow is the presence of a companion accreting gas from the AGB star. Indeed,
$\pi^1$ Gru and W Aql (\cite{Hoai2015}) are both S type AGB binaries with a spatially
resolved companion. Unfortunately, in both cases, the orbital period is too long to allow
for an unambiguous evaluation of the inclination of the orbit on the plane of the sky,
making it difficult to assess the role played by the companion in shaping the CSE.

The above comments should not be taken as general properties of AGB stars, but
as applying to only a family of such stars. The diversity of the features observed on stars
evolving on the asymptotic giant branch (\cite{De2010}) must always be kept in
mind before attempting at stating general laws. Analyses of observations of the gas CSE
of AGB stars provide invaluable information on their morphology and kinematics, as has
been amply demonstrated recently using millimeter and submillimeter interferometer
arrays. The present state of the art of our understanding of the mass loss mechanism
strongly suggests that the unprecedented spatial resolution and sensitivity made available
by ALMA will open a new era in this field of astrophysics. Typical AGB stars, such as
those mentioned above, are ideal targets for future multiline ALMA observations.

\section*{ACKNOWLEDGEMENTS}
We are indebted and very grateful to the ALMA partnership, who are making their data
available to the public after a one year period of exclusive property, an initiative that
means invaluable support and encouragement for Vietnamese astrophysics. We
particularly acknowledge friendly and patient support from the staff of the ALMA
Helpdesk. We express our deep gratitude to Professors Nguyen Quang Rieu and Thibaut
Le Bertre for having introduced us to radio astronomy and to the physics of evolved stars.
Financial support is acknowledged from the Vietnam National Satellite Centre
(VNSC/VAST), the NAFOSTED funding agency, the World Laboratory, the Odon Vallet
Foundation and the Rencontres du Viet Nam.

% UNCOMMENT THE LINES BELOW IF YOU WISH TO USE BIBTEX
%\bibliographystyle{apj}
%\bibliography{yourbibfile}
%\begin{thebibliography}{}

\label{lastpage}

\end{document}